\documentclass[aps,prl,twocolumn,showpacs,superscriptaddress,groupedaddress]{revtex4}
\usepackage[pdftex]{graphicx}
\usepackage{float}
\usepackage{dcolumn}
\usepackage{bm}
\usepackage{amssymb}
\usepackage{rotating}
\begin{document}
\widetext
\title{Soliton Turbulence in Shallow Water Ocean Surface Waves}
\author{Andrea Costa} 
\affiliation{Aix-Marseille Universit\'e, CNRS/INSU, IRD, MIO, UM 110, 13288, Marseille, Cedex 9, France;
Universit\'e de Toulon, CNRS/INSU, IRD, MIO, UM 110, 83957, La Garde, France }
\email[]{andrea.costa@univ-amu.fr}
\author{Alfred R. Osborne} 
\affiliation{Nonlinear Waves Research Corporation, Arlington, VA 22203, U. S. A. } 
\email[Corresponding author: ]{al.osborne@gmail.com}
\author{Donald T. Resio} 
\affiliation{Department of Ocean Engineering, University of North Florida, Jacksonville, FL 32224-7699, U. S. A. } 
\author{Silvia Alessio} 
\affiliation{Dipartimento di Fisica, Universit\`{a} di Torino, Torino 10125, Italy } 
\author{Elisabetta Chriv\`{i}} 
\affiliation{Dipartimento di Fisica, Universit\`{a} di Torino, Torino 10125, Italy } 
\author{Enrica Saggese} 
\affiliation{Universit\'{e} Nice Sophia Antipolis, LPMC, UMR 7336, 06100 Nice, France} 
\author{Katinka Bellomo} 
\affiliation{Rosenstiel School of Marine and Atmospheric Science, University of Miami, Miami, FL 33149, U. S. A.} 
\author{Chuck E. Long} 
\affiliation{U. S. Army Engineer Research and Development Center, Vicksburg, Mississippi, U. S. A., retired.}
\date{\today}
\begin{abstract}
We analyze shallow water wind  waves in Currituck Sound, North Carolina and experimentally confirm, for the first time, the presence of \textit{soliton turbulence} in ocean waves. Soliton turbulence is an exotic form of nonlinear wave motion where low frequency energy may also be viewed as a \textit{dense soliton gas}, described theoretically by the soliton limit of the Korteweg-deVries (KdV) equation, a \textit{completely integrable soliton system}: Hence the phrase ``soliton turbulence" is synonymous with ``integrable soliton turbulence." For periodic/quasiperiodic boundary conditions the \textit{ergodic solutions} of KdV are exactly solvable by \textit{finite gap theory} (FGT), the basis of our data analysis. We find that  large amplitude measured wave trains  near the energetic peak of a storm have low frequency power spectra that behave as $\sim\omega^{-1}$.  We use the linear Fourier transform to estimate this power law from the power spectrum and to filter \textit{densely packed soliton wave trains} from the data. We apply FGT to determine the \textit{soliton spectrum} and find that the low frequency $\sim\omega^{-1}$ region is \textit{soliton dominated}. The solitons have \textit{random FGT phases}, a \textit{soliton random phase approximation}, which supports our interpretation of the data as soliton turbulence. From the \textit{probability density of the solitons} we are able to demonstrate that the solitons are \textit{dense in time} and \textit{highly non Gaussian}. 
\end{abstract}
\maketitle

\begin{figure*}
\includegraphics[width=17.9cm, height=4.0cm]{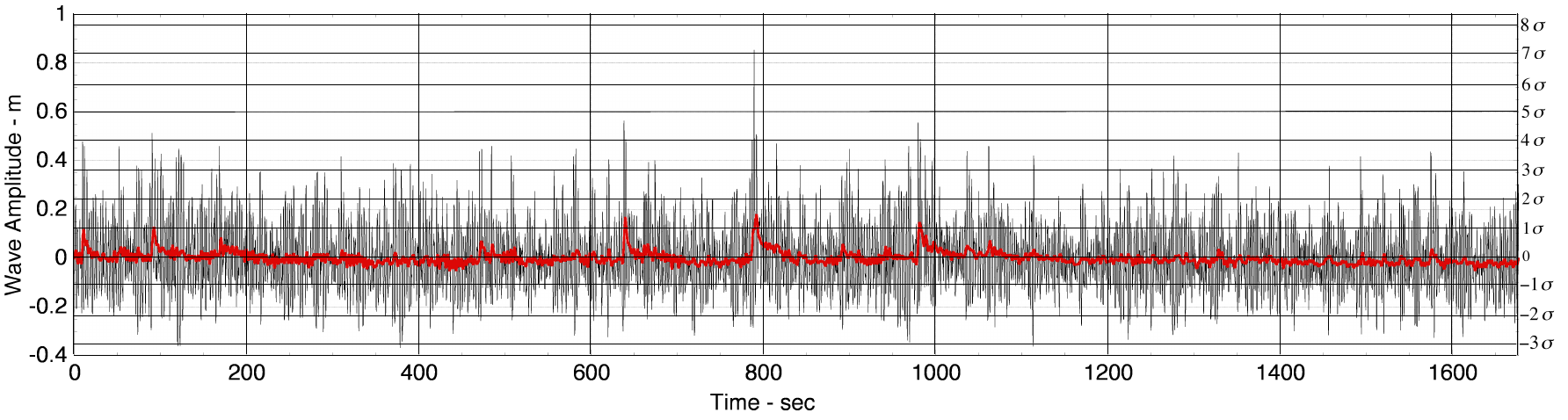}
\caption{Measured surface wave time series of 8192 points (27.96 minutes, sampling interval $0.2048$ $s$, black curve) from Currituck Sound beginning at 21:00h  on 4 February 2002. The significant wave height was 0.52 m in a depth of 2.63 m. The red curve is the low frequency soliton signal obtained by low pass filtering the (black) measured time series.}
\end{figure*} 
\begin{figure}
\includegraphics[width=8.5cm, height=7.1cm]{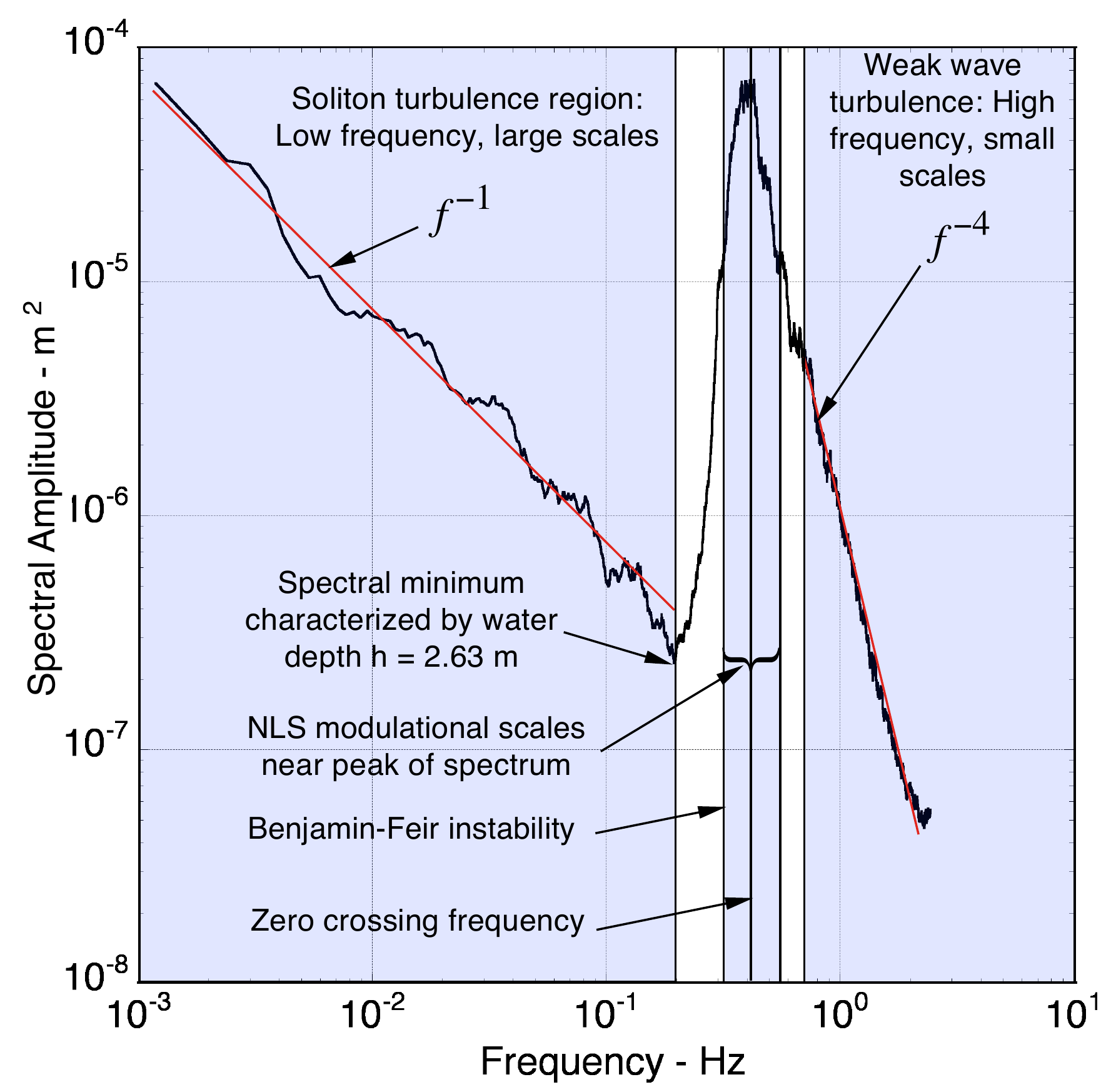}
\caption{Power spectrum of the measured time series in  FIG. 1. Validity intervals for KdV ($f<0.22$ Hz) and NLS ($0.34$ Hz $<f<0.56$ Hz) are shown. Exact power laws (red lines) are shown in the low-frequency soliton turbulent region ($\sim f^{-1}$) and high-frequency cascade region ($\sim f^{-4}$).}
\end{figure}
\begin{figure}[!h]
\includegraphics[width=8.5cm, height=7.1cm]{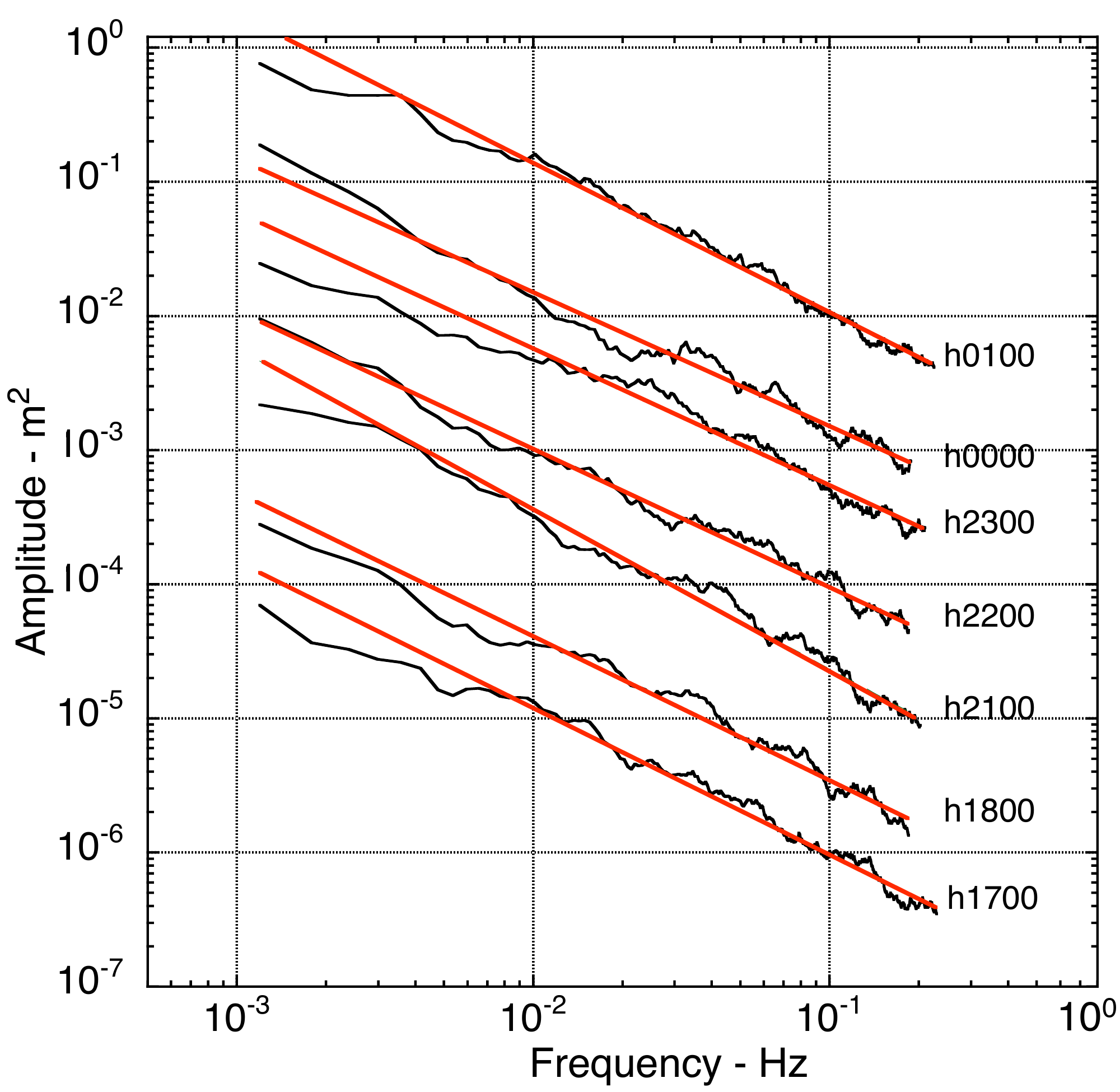}
\caption{Least squares fit spectral power laws of the experimentally determined solitonic wave trains at different hours during a Currituck Sound storm on 04/02/2002. The spectra have been vertically shifted for clarity.}
\end{figure}
\begin{figure}[!h]
\includegraphics[width=8.5cm, height=4.7cm]{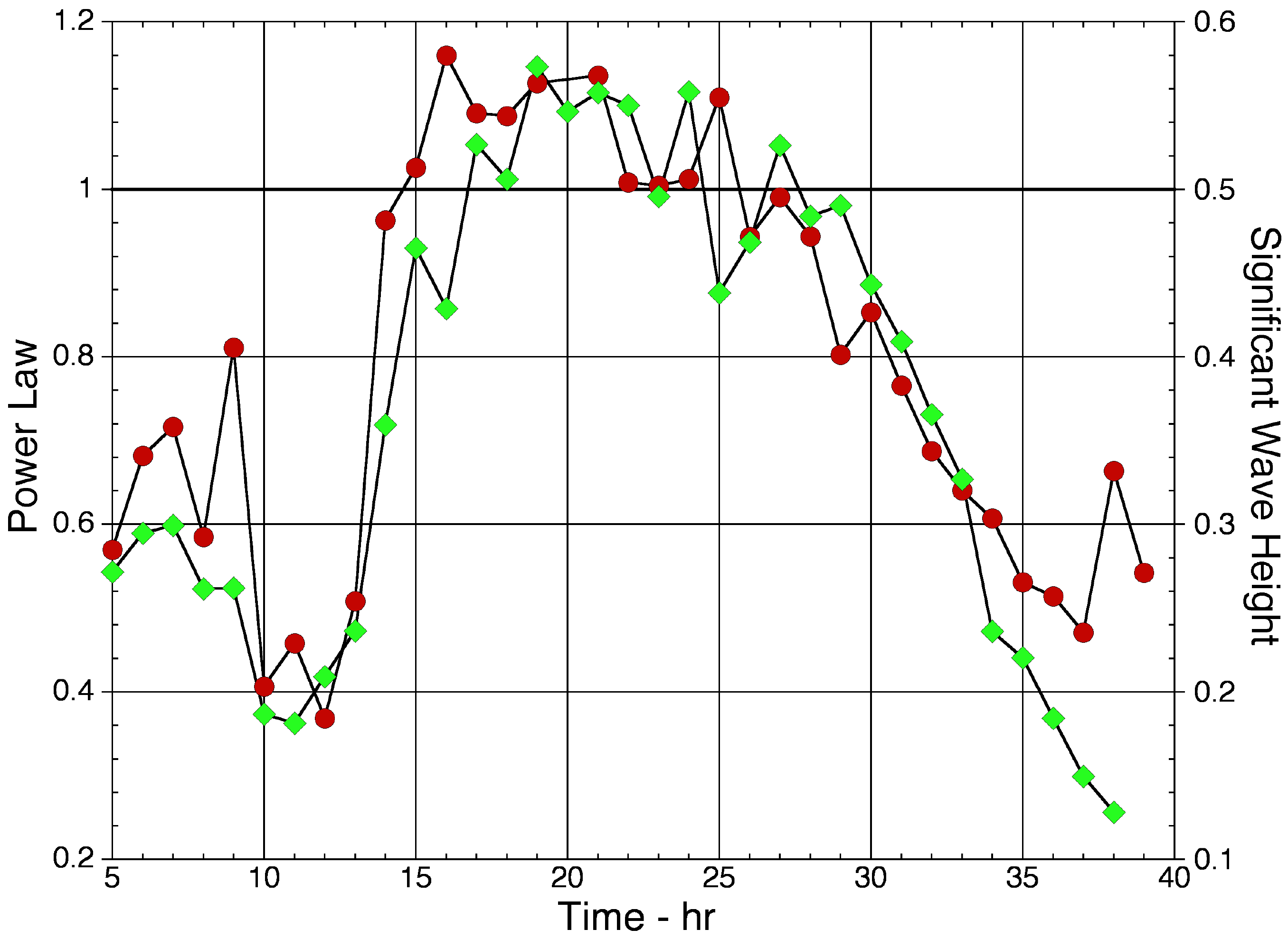}
\caption{Significant wave height $H_s$ during the Currituck Sound storm (green diamonds) and slope of power law spectra $\gamma$  (red circles) versus time during the Currituck Sound storm.} 
\end{figure}
\begin{figure*}
\includegraphics[width=17.4cm, height=4.65cm]{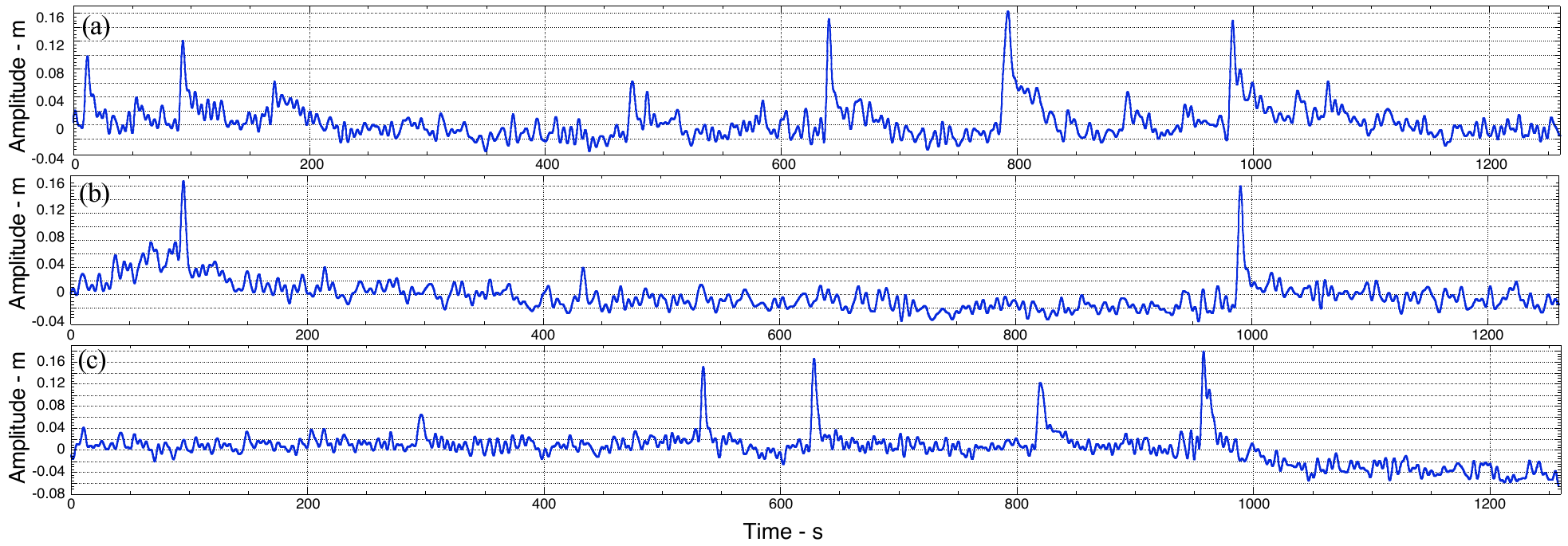}
\caption{Turbulent soliton wave trains computed by low-pass filtering the measured wave data as discussed in the text. The results were obtained during the storm of 4 February 2002 (a), (b) and (c). We have verified with FGT that the peaks in these time series are solitons and are governed by a low frequency power law as shown in FIGs. 2, 3.}
\end{figure*}
\begin{figure*}
\includegraphics[width=17.9cm, height=5.0cm]{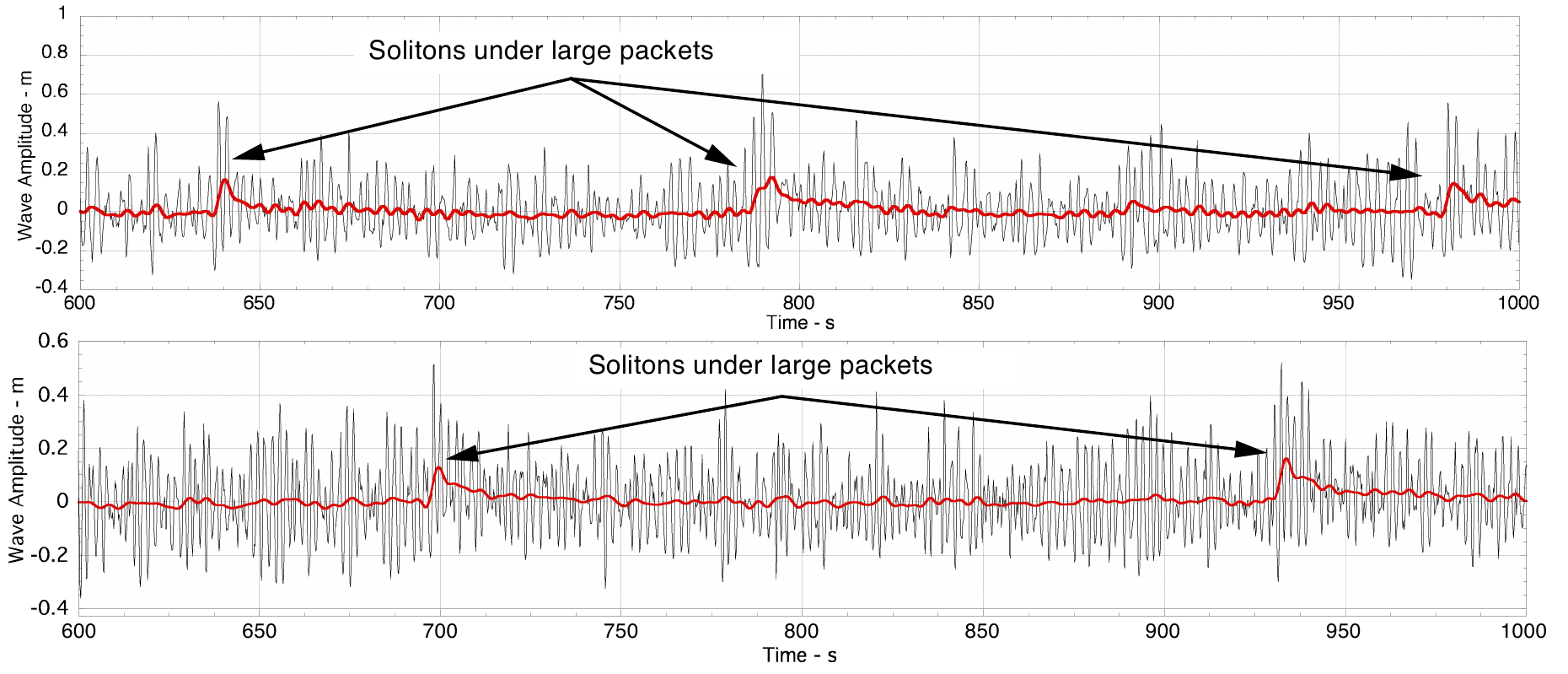}
\caption{Two measured wave trains  (black) together with the underlying soliton trains obtained by low pass filtering of the data (red). The results show how large solitons tend to occur under large packets.}
\end{figure*}
\begin{figure}
\includegraphics[width=8.6cm, height=4.8cm]{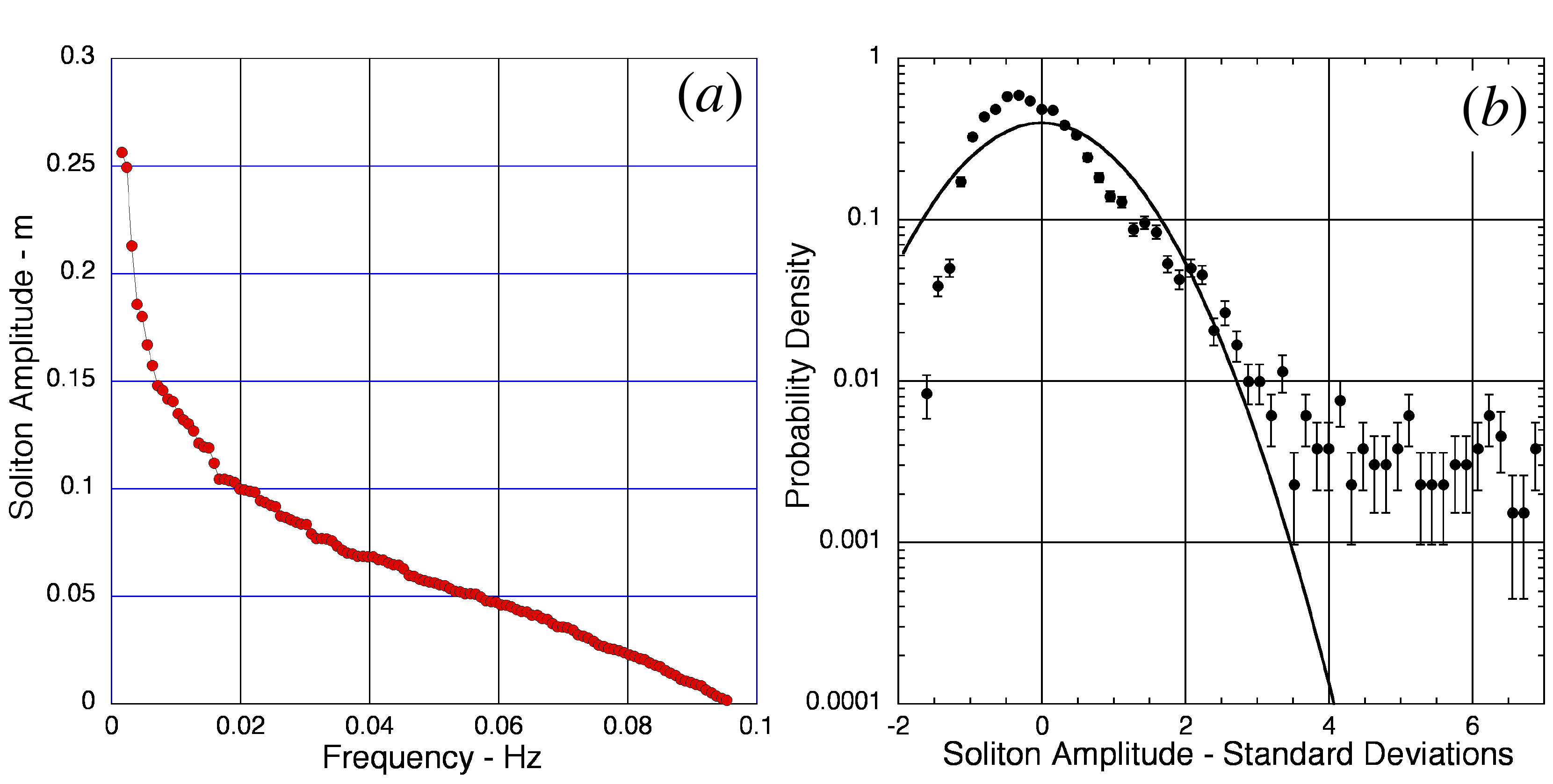}
\caption{FGT Soliton spectrum of the measured wave train in  FIG. 1 (a). Histogram of soliton amplitudes from  FIG. 5  (b). The highly non Gaussian nature of the solitons is clear.}
\end{figure}
\indent
The physical basis of \textit{weak wave turbulence} was developed by Zakharov and Filonenko \cite{zakharovfilonenko_WWT_4_1967}. They investigated the theoretical power spectrum for ocean surface waves and demonstrated that in deep water the direct cascade of energy - from the spectral peak to higher frequencies in the spectral tail - should be of the form $E(\omega)\sim\omega^{-4}$. This theoretical result was confirmed in subsequent work \cite{zakharov_WWT_1967_1}, \cite{zakharov_WWT_1999_1} in which the power law was found to be an exact solution of the kinetic equation for the waves. The expansion used in this computation is only up to the third order in wave steepness and thus the theory is referred to as ``weak turbulence." Both numerical and experimental confirmations have been found \cite{ResioPerrie, LongResio, Pushkarev, Toba1, Toba2, Dyachenko, Pelinovsky}. Zakharov \cite{zakharov_WWT_1967_1, zakharov_WWT_1999_1} also found the $Kolmogorov$ $spectrum$ for \textit{shallow water} weak wave turbulence. The inverse cascade of \textit{wave action} to large scales/small frequency is a power law: $I_{\omega}  \sim Q^{1/3} \omega^{-1}$ where $Q$ is the flux of action.\\ 
\indent
The theory of \textit{integrable soliton turbulence}, as used here to analyze ocean wave data, is based on the discovery of \textit{complete integrability} for the Korteweg-deVries (KdV) equation: 
\begin{equation} \label{kdv}
\eta_t+c_0\eta_x+\alpha\eta\eta_x+\beta\eta_{xxx}=0
\end{equation}
($c_o  = \sqrt {gh} $, $\alpha  = 3c_o /2h$, $\beta  = c_o h^2 /6$, for $h$ the water depth, $g$ the gravitational acceleration), valid for small but finite amplitude, long waves in shallow water. KdV is integrated by the Inverse Scattering Transform (IST)  on the infinite line \cite{GGKM}. Zakharov has studied this shallow water case \cite{zakharov_IST_1971, zakharov_IST_2009} for integrable turbulence for a \textit{rarified soliton gas}. He derived a \textit{soliton-gas kinetic equation} for the KdV equation using the IST. More recently the  kinetic equation for a \textit{dense soliton gas} for integrable nonlinear wave equations has been found by El and Kamchatnov \cite{El_2005} by taking the thermodynamic limit of the Whitham equations to obtain a nonlinear integro-differential equation for the spectral measure. This result generalizes Zakharov's case for a rarified soliton gas.\\
\indent
The other equation we refer to herein is the nonlinear Schr\"odinger (NLS) equation which describes nonlinear wave packet dynamics  
\begin{equation} \label{NLS}
i(\psi_t+C_g\psi_x)+\mu\psi_{xx}+\nu|\psi|^2\psi=0
 \end{equation}
($C_g$, $\mu$ and $\nu$ are depth dependent constants \cite{Mei_1983}). The NLS equation is approximately valid in a narrow band about the spectral peak. We use NLS here mainly to ensure the separation of long wave (KdV) and short wave (NLS) scales as discussed below in the data analysis.\\
\indent
Herein we test to ensure that the measured time series are stationary and ergodic, a standard procedure for the analysis of ocean waves. The fast Fourier transform (FFT, a periodic algorithm) is the most often used method for data analysis. Likewise finite gap theory \cite{Bel_1994} (nonlinear Fourier analysis for KdV which is also periodic) is used to analyze and interpret the measured Currituck Sound data using the methods of  \cite{Osborne_2010}. This means that we are able to deal, from a theoretical and data analysis point of view, with the \textit{densely packed solitons} found in the  data. Herein, our use of the term \textit{soliton turbulence}  is synonymous with \textit{integrable soliton turbulence} as discussed in the theoretical literature \cite{zakharov_IST_1971, zakharov_IST_2009, El_2005}. The $\omega^{-1}$ theoretical power law of Zakharov \cite{zakharov_WWT_1967_1, zakharov_WWT_1999_1} for shallow water weak wave turbulence is \textit{not} applicable to \textit{high density soliton interactions} with \textit{strongly non Gaussian} behavior as addressed experimentally herein.\\
\indent
A confirmation of the theoretical behavior of soliton dynamics of integrable soliton gases came from numerical simulations using FGT \cite{Osborne_1993}. The method was applied to construct \textit{realizations} of KdV random processes with a power law spectrum $k^{-\gamma}$ and \textit{uniformly distributed FGT phases}. These highly nonlinear cases consisted of energetic, densely packed solitons in low-level radiation.\\
\indent
Direct experimental verification of soliton turbulence in the ocean has remained unconfirmed for over four decades. One obstacle has been the impossibility of distinguishing by eye solitons from the \textit{large radiative (wind) waves} in experimental data. This difficulty was overcome in \cite{Osborne_1991} using a nonlinear filtering technique - based on FGT for KdV - to extract solitons from surface wave data obtained in the Adriatic Sea.\\
\indent
In the present paper we analyze data measured by Long and Resio in Currituck Sound, North Carolina \cite{LongResio}. FIG. 1 shows a measured amplitude time series, whose power spectrum is given in  FIG. 2. An important characteristic of the Currituck Sound data is the small depth ($h$=2.63 $m$) in which the probes were positioned. This particular water depth allowed the simultaneous measurement of the \textit{spectrally well-separated dynamics} of both KdV (shallow water wave dynamics) and NLS (variable depth dynamics  centered about a narrow spectral peak)  in the same data set, resulting in a power spectrum which divides high frequency and low frequency behaviors by a \textit{low energy spectral minimum} parametrically characterized by the depth $h$. The mean of the frequencies at the \textit{minima in the measured spectra}, in all data sets analyzed is $f_{min} \simeq 0.22$  Hz,  corresponding to a value of $kh\simeq0.80$, where $\omega^2 = g k \tanh(k h)$ is the linear dispersion relation, $f=\omega/2 \pi$  (see FIG. 2). Thus, the experimental set up is quite unique, with well separated KdV and NLS behaviors. This contrasts to \cite{Osborne_1991} where no NLS regime occurred.\\
\indent
The high-frequency cascade range of the wind-wave spectrum, $f >0.7$ Hz, was found to have a power-law $\sim f^{-4}$ (FIG. 2), in agreement with \cite{ResioPerrie, LongResio, Toba1, Toba2}. The low-frequency spectra were also found to be approximated by a power law $\sim f^{-\gamma}$ during the full 34 hours of the storm, FIG. 2. FIG. 3 shows the least squares fits of many of these low frequency power law spectra found during the peak of the storm. In FIG. 4 we graph the significant wave height $H_s$ and the low-frequency slope $\gamma$ as a function of time over the period of the storm. Near the peak of the storm $H_s$ averaged 0.496 $\pm$ 0.060 $m$ and $\gamma$ averaged 1.043 $\pm$ 0.074.\\
\indent
Several well-defined, large amplitude solitons were found in the present study. In  FIG. 5 we show three turbulent soliton trains in the absence of the background radiation and in  FIG. 6 we show several soliton trains beneath the measured surface waves. The tendency for the largest solitons to occur beneath large wave packets is clear. Thus our data set vastly extends on previous results from  \cite{Osborne_1991} where this effect was first seen.\\
 \indent
In order to characterize the measured soliton wave trains we have: (1) Extracted the long wave, low frequency part of the spectrum from the measured data to see the soliton turbulence in the absence of the radiation modes (FIGS. 1, 5 and 6). (2) Computed the power spectrum in order to obtain the \textit{spectral slope} $\gamma$ (FIGs. 3, 4). (3) Determined that these low frequency Fourier power spectral components are solitons using FGT (FIGs. 5, 6). (4) Computed the \textit{soliton spectrum} (FGT) and the \textit{probability density of solitons} (FIG.7). In the \textit{first method} (1) we low pass filtered the measured time series using the fast Fourier Transform and FGT. Because of the well-separated scales in the spectral domain, these results are comparable and support the soliton interpretation of the data. In the \textit{second method} (2) we use the Fourier transform to obtain the power spectrum to estimate the slope of the power law $\gamma$ (FIG. 3). In the \textit{third and forth methods} (3), (4) we apply FGT to compute the \textit{nonlinear spectrum} to determine whether the power law spectrum computed from the linear Fourier transform arises strictly from solitons: For each \textit{elliptic modulus} near 1 we have a soliton component. FGT demonstrates that the soliton modes saturate the low-frequency part of the power spectrum, spanning the region of the $\sim f^{-1}$ power-law.\\
\indent
The nonlinear physics corresponds to a \textit{dense soliton gas} as seen in the \textit{nonlinear spectrum} and \textit{probability density function} (FIG. 7). For each of the 14 time series near the storm peak there are about 120 solitons that appear in the region of the low frequency power spectra characterized by a power law $\gamma \sim $ 1.043 $\pm$ 0.074. The average full width at half maximum of each soliton is about  10.5 sec (1258 sec/120 solitons): roughly half of the solitons are smaller (and broader) than the average soliton (6.3 cm height) and are therefore more densely packed than the average, while the remaining half of the solitons are larger (and more narrow) than the average and are thus less dense and easily seen as the largest solitons in FIGS. 1, 4 and 5. We also find that the \textit{FGT phases} of the solitons are random numbers on $(0,2\pi )$, thus connecting integrable FGT with a statistical description of the data, the \textit{solitonic random phase approximation} of FGT \cite{Osborne_1993}: Our data is described by soliton FGT modes with random phases, which is soliton turbulence, the random soliton limit of KdV.\\
\indent
Reasons why the Currituck Sound experiment has been able to successfully measure soliton turbulence include: (1) The shallow water depth allows for the generation of long wave solitonic components. (2) The \textit{particular depth} of 2.6 m divides the low frequency KdV region of the spectrum from the high frequency NLS region. (3) Large wave conditions occurred at the peak of the storm on 5/2/2002, thus providing a large range of nonlinear frequency scale interactions in the spectrum. (4) Use of FGT allows us to determine the presence of soliton turbulence in the spectrum of the data.\\
\indent
We acknowledge partial support from the Army Corp of Engineers and the Office of Naval Research.\\

\end{document}